\documentclass{article}
\input{tcilatex}

\begin{document}

\title{{\LARGE Regge Amplitudes and Final State Phases in the Decays }$B$%
{\LARGE \ }$\longrightarrow ${\LARGE \ }$D\pi $}
\author{{\large Lincoln Wolfenstein} \\
%EndAName
{\small Department of Physics, Carnegie Mellon University, Pittsburgh, PA
15213}}
\maketitle

\begin{abstract}
The scattering amplitude of $D\pi $ at the energy of the $B$ mass can be
calculated using Regge theory. \ Recent papers have used this to calculate
the final state strong phases in the decays $B\longrightarrow D\pi .$ \ It
is argued that while the Regge amplitude can yield an absorption correction
to the decay rate, it is not useful for determining the strong phase.
\end{abstract}

\qquad \vspace{1in}

A number of recent papers have used a Regge analysis to determine the strong
phases in the amplitude of $B$ decay $\longrightarrow $ $D\pi $ (1-3). Here
we discuss the difficulties with these analyses...

The starting point is the generalized Watson theorem based on CPT and
unitarity

\begin{equation}
A_{f}=\sum_{i}\left( S\right) _{fi}^{\frac{1}{2}}\ M_{i}\quad
\end{equation}%
where $A_{f}$ is the decay amplitude to state $f$, $M_{i}$ is the weak decay
amplitude in the absence of final-state scattering and $S$ is the scattering 
$S$ matrix at an energy equal to the $B$ mass. \ This scattering is
primarily due to strong interactions. The sum is over all states to which $B$
may decay.

In these papers the sum is truncated at the first term, the "elastic
scattering" term. As discussed later, this truncation is completely
unjustified, but first we pursue this approximation, yielding $A=(S)^{\frac{1%
}{2}}\ M$ where we have omitted the subscript $f$ so that $S=S_{ff}$ and we
need only consider $s$-wave scattering. $S$ is related to the scattering
amplitude $F$ by

\begin{equation}
S=1+2iF
\end{equation}

For the case of $B$ $\longrightarrow $ $D\pi $ there are two final states $%
D^{+}\pi ^{-}$ and $D^{0}\pi ^{0}$; \medskip however using the isospin
analysis we can consider separately the $I=\dfrac{3}{2}$ and $I=\dfrac{1}{2}$
states as the single state $f$.

For the case of truly elastic scattering Eq. (2) gives

\begin{equation}
S=1+2i\left( \sin \delta e^{i\delta }\right) =e^{2i\delta }
\end{equation}%
so that the phase of $F$ gives directly the phase of $\delta $ of $S^{\frac{1%
}{2}}$, but this is not true in the case of $B\longrightarrow D\pi $, where
the scattering is primarily inelastic. In Refs. 2 and 3 it seems that they
in fact do assume the phase of the Regge amplitude gives the final state
phase (see just below Eq 26 of Ref.3), which is clearly wrong. If one only
considers the Pomeron trajectory the amplitude $F$ is purely imaginary so
that, from Eq 2, $S$ is real and there is no strong phase. The Pomeron
represents the effect of diffractive scattering that scatters out from $D\pi 
$ but not back into it.

In the case that the scattering is not purely elastic the element $S_{ff}=S$
of the $S$-matrix can be written

\begin{equation}
S=\eta e^{2i\delta }
\end{equation}%
It should be noted that putting this into Eq (1) gives a factor $\eta ^{%
\frac{1}{2}}$ in addition to the strong phase. This may be considered as an
"absorption correction" to the simple calculation of $M$ due to the
reduction of $M$ from rescattering to other states. Such a correction was
introduced by Sopkovich \cite{Sopkovich} in the context of the
meson-exchange calculation of a scattering amplitude and applied extensively
by Gottfried and Jackson in this context \cite{Gottfried}.

In the case of $B$ decays if $M$ is assumed to be calculated accurately in
the absence of strong rescattering one should multiply the result by the
square root of $\eta $. In practice this is not usually done and presumably
it is absorbed in the determination of some of the parameters entering the
calculation of $M$. If the Pomeron dominates the inelastic scattering then
using it in Eq.2 may give a good value of $\eta $ even though it is
essentially irrelevant for determining the strong phase.

In the case of eq(4) the phase $\phi $ of $F$ is related to the phase $%
\delta $ of $S$ by

\begin{eqnarray}
\tan \phi &=&\tan \delta +A(\tan \delta +\cot \delta ) \\
A &=&\frac{(1-\eta )}{2\eta }  \nonumber
\end{eqnarray}

Fayazuddin \cite{Fayazuddin2} uses the correct relation between the phase of
the $S$ matrix and the phase of the Regge amplitude as illustrated in Table
3 of that paper. In the examples shown the main effect of the Regge
amplitude is to give a value of $\eta $ around 0.7 while the phase delta is
quite small.

In Ref (1) Fayazuddin applies this to the decay $B\longrightarrow D\pi $
with results that appear to agree with experiment. However there is still
the serious problem of the truncation of Eq. (1). In the Regge analysis the
non-zero phase of $\delta $ arises first via $\rho $ exchange which gives a
real contribution to the elastic scattering amplitude $F$.

\bigskip However $\rho $ exchange also gives $D^{\ast }\pi \longrightarrow
D\pi $ scattering and since $M$ for $B\longrightarrow D^{\ast }\pi $ is much
the same as $M$ for $D\pi $ this should double the phase at least in the
small phase limit. There also are contributions from the $D^{\ast }\rho $
state although this requires $\pi $ exchange.

In conclusion the $D\pi $ cross-section consists of highly inelastic
scattering such as that described by the Pomeron which primarily determines
the parameter $\eta $ in the $S$ matrix. There is also two body to two body
scattering such as that described by $\rho $ exchange. This includes the
elastic scattering which contributes to the strong phase $\delta $ but also
scattering into the $D\pi $ state from other final states which may be least
important as the elastic in determining $\delta $.

This research was supported in part by the U.S. Department of Energy grant
No. DE-FG02-91ER40682.

I am indebted to discussions with Mahiko Suzuki and the hospitality of the
University of California and the Lawrence Berkeley Laboratory.

\end{document}